\documentclass{iucrjournals}

\nolinenumbers

\RequirePackage{hyperref}

\usepackage[final]{spin-textures}
\RequirePackage{orcidlink}

\graphicspath{{Images}}

\title{Time-resolved X-ray Magnetic Circular Dichroism Ptychography at the HERMES beamline of synchrotron SOLEIL}
\author[a,b,c]{Luc\'\i a G\'omez-Cruz\IUCrOrcidlink{0009-0006-9777-429X}}%
\author[a,d,e]{Niklas Martin\IUCrOrcidlink{0009-0008-5965-518X}}
\author[b,c]{Lucas Perez\IUCrOrcidlink{0000-0001-9470-7987}}
\author[d]{Christophe Thirion\IUCrOrcidlink{0000-0001-6400-7971}}
\author[a]{Aurélien Masseboeuf\IUCrOrcidlink{0000-0003-4239-1313}}
\author[a]{Olivier Fruchart\IUCrCemaillink{olivier.fuchart@cea.fr}\IUCrOrcidlink{0000-0001-7717-5229}}
\author[f]{Rachid Belkhou\IUCrOrcidlink{0000-0002-2218-7481}}%

\affil[a]{Univ. Grenoble Alpes, CNRS, CEA, SPINTEC, Grenoble, France}
\affil[b]{IMDEA Nanociencia, Campus de Cantoblanco, 28049 Madrid, Spain}
\affil[c]{Dpto. de F\'\i sica de Materiales, Universidad Complutense de Madrid, 28040 Madrid, Spain}
\affil[d]{Univ. Grenoble Alpes, CNRS, Institut Néel, Grenoble, France}
\affil[e]{Karlsruhe Institute of Technology, 76131 Karlsruhe, Germany}
\affil[f]{Synchrotron SOLEIL, l’Orme des Merisiers, Saint-Aubin, FR-91192 Gif-sur-Yvette Cedex, France}

\begin{document}
\doclicenseThis
\maketitle

%\preprint{AIP/123-QED}

\date{\today}

\begin{synopsis}
The HERMES beamline at SOLEIL synchrotron now offers time resolution for ptychography soft-X-ray magnetic imaging. This opens opportunities for addressing unexplored aspects of magnetization dynamics at the \qty{10}{\nano\meter} length scale.
\end{synopsis}

\begin{abstract}
We report the implementation of time resolution in soft-X-ray ptychography magnetic imaging at the HERMES beamline of synchrotron SOLEIL. We demonstrated the monitoring of domain-wall dynamics and magnetization switching with sub-\qty{100}{\pico\second} resolution, spin-wave emission and propagation, and magnetic resonances, up to several \si{\giga\hertz}. Owing to the capability of ptychography to reach or exceed \qty{10}{\nano\meter} spatial resolution, of acute relevance for nanomagnetism as comparable with the dipolar exchange length, we believe that time-resolved ptychography magnetic imaging will prove to be very valuable in the coming years. Partial filling modes of synchrotrons are particularly relevant for the physical and heat-management time scales of spintronics.
\end{abstract}


\section{Context}

\subsection{Soft X-ray ptychography}

\subsection{Needs and assets for magnetic imaging}

The primary criteria for magnetic imaging is the ability to resolve magnetization distributions. Magnetic materials have a number of typical length scales whose relevance depends on the situation. The most prominent length scales are the anisotropy exchange length $\AnisotropyExchangeLength=\sqrt{A/K_\mathrm{u}}$ and the dipolar exchange length $\DipolarExchangeLength=\sqrt{2A/\muZero\Ms^2}$, with $K_\mathrm{u}$ an anisotropy coefficient, $A$ the exchange stiffness, and $\Ms$ the spontaneous magnetization\cite{bib-HUB1998b}. The former length scale, also called Bloch parameter, results from the competition of magnetocrystalline anisotropy and exchange. Its value varies over many orders of magnitude depending on the material, reflecting mostly the large spread of values of magnetic anisotropy. It ranges from typically one nanometer in hard-magnetic materials to hundreds of nanometers in soft-magnetic ones. Comparatively, the exchange stiffness and magnetization do not vary much among common 3d ferromagnets and their alloys, of the order of $\qty{1E-11}{\joule\per\meter}$  and $\qty{1E6}{A/m}$, respectively. Consequently, the dipolar exchange length has similar values in these materials, typically in the range \qtyrange{3}{7}{\nano\meter}. Therefore, magnetic microscopies with spatial resolution \qty{10}{\nano\meter} or smaller may be desirable in situations where magnetostatics and exchange compete, such as for vortex cores\cite{bib-SHI2000b}, or in high-anisotropy materials, such as ultrathin films with perpendicular magnetization resulting from interfacial magnetic anisotropy\cite{bib-JOH1996}. Other material parameters can influence characteristic magnetic length scales, such as the Dzyaloshinskii-Moriya interaction for skyrmions\cite{bib-BER2018b}, applied magnetic field, or sample size.

Another criteria of importance for magnetic microscopy is sensitivity, in the context that most nanomagnetism and spintronic effects arise from interfaces and thus require layer thicknesses in the nanometer range. Conversely, the probing depth matters for thicker films or surfaces of bulk samples, and is also gaining importance in the context of the rise of three-dimensional~(3D) nanomagnetism\cite{bib-FRU2017b,bib-MAK2022}. To achieve depth sensitivity, a first aspect is to be able to probe the entire volume of a sample, which requires long-range interactions for techniques measuring stray fields, or depth penetration and sufficient flux for imaging with waves or particles. Many structures in 3D nanomagnetism have a thickness of the order of \qtyrange{100}{200}{\nano\meter}. In some situations, it is also important not only to get a two-dimensional~(2D) map such as a surface or projection, but also the depth variation of magnetization. Techniques such as tomography or similar are then desirable.

Finally, while experimental techniques are scalar by nature as they measure the magnitude of a given quantity, one may desire vectorial information. Within 2D systems, a typical example is skyrmions, whose deterministic chirality is the signature making them distinct from usual magnetic bubbles\cite{bib-FER2017}. In 3D objects, the variety of predicted or proven magnetization textures is increasing rapidly, all involving a  full $4\pi$ span of solid  angle of magnetization: skyrmion tubes, bobbers and cocoons, hopfions, vorticity strings etc.\cite{bib-MAK2022}

\subsection{Needs and assets for time-resolved imaging}

Magnetization dynamics in magnetic materials results from various physical effects and covers many orders of magnitude: thermally-activated stochastic events from years to the $\si{\nano\second}$\cite{bib-NEE1949}, gyration of vortices\cite{bib-BUE2004,bib-HRK2015} and other magnetization textures and spin-torque nano-oscillators in the range of tens to hundreds of $\si{\mega\hertz}$\cite{bib-DEV2007}, ferromagnetic resonance in ferromagnets in the $\si{\giga\hertz}$ range. Faster phenomena include ferromagnetic resonance in antiferromagnets, liable to reach several hundreds of $\si{\giga\hertz}$, or the recent field  of $\si{\tera\hertz}$  spintronics based on the spin-to-charge conversion arising from light-induced magnetization dynamics\cite{bib-KAM2013}, itself down to the picosecond time scale or lower\cite{bib-LAM2014}.

While the large flux of X-rays at free-electron lasers has made possible the emergence of single-shot microscopy\cite{bib-GUT2010}, the vast majority of techniques requires averaging and a stroboscopic scheme to reconstruct a time structure. This sets a number of limitations. First, on the side of the physical device, the dynamic processes must be reproducible to a large extent, up to billions of times, so that the averaged measurement reflects the underlying phenomenon. Similarly,  not only does the response of the system need to be reproducible, but one must ensure that the initial state is the same for each individual measurement, which can set additional constraints on the instrumentation. Finally, in the case of spintronic effects, the exciting current pattern must be fed continuously to the device, which can raise issues of temperature management and breakdown.

\subsection{Soft X-ray imaging}

In the context summarized above, soft X-ray microscopy has become an essential tool, owing to its unique combination of elemental selectivity, sensitivity to electronic structure and ability to probe magnetic properties through resonant interactions. In particular, the use of soft X-rays at the absorption edges of transition metals provides access to element-specific information on chemical bonding, oxidation states and magnetic ordering. These capabilities make soft X-ray microscopy particularly relevant to the study of functional materials, magnetic heterostructures and nanostructured systems.

Transmission X-ray microscopy~(TXM)\cite{bib-FIS1997b} and its scanning version, STXM\cite{bib-KOR2000b}, have gained an increasing importance over the past twenty years\cite{bib-FIS2001,bib-FIS2015}. They both make use of Fresnel zone plates acting as lenses to manipulate the X-ray beam, and operate in transmission through the sample. TXM is a full-field imaging technique, with a first zone plate before the sample focusing the beam over the area of interest, while a second zone plate after the sample builds the image in the plane of a 2D detector. STXM makes use of a single zone plate focusing the beam at its maximum, while the transmitted light is monitored on a photodiode. The spatial resolution of soft X-ray (S)TXM is of the order of \qty{25}{\nano\meter}, determined by the zone plate(s). When combined with X-ray Magnetic Circular Dichroism~(XMCD), (S)TXM provides information about the component of magnetization parallel to the X-ray beam. X-ray Magnetic Linear Dichroism~(XMLD) allows to image some antiferromagnets\cite{bib-STO1998}.

In this context, soft X-ray ptychography has emerged as a powerful approach to quantitative nanoscale imaging\cite{bib-SHA2014,bib-HIT2015,bib-PFE2017}. Combining scanning transmission microscopy with coherent diffraction imaging, ptychography relies on the acquisition of far-field diffraction patterns from overlapping illuminated regions of a specimen. Iterative phase-retrieval algorithms are subsequently used to reconstruct the complex transmission function of the sample, providing both amplitude and phase information\cite{bib-BUN2008,bib-DIE2010}. Unlike conventional lens-based microscopy, the spatial resolution is not directly limited by the imaging optics, but is primarily determined by the maximum scattering angle collected by the detector, together with the coherence and stability of the experimental system\cite{bib-PFE2017}. Soft X-ray ptychography presents a number of assets for magnetic microscopy, when combined with XMCD or XMLD. First, the spatial resolution has been proven to be \qty{10}{nm}, and is expected to further improve thanks to higher coherence at fourth-generation synchrotrons, and cameras with a larger number of pixels. It is also pretty much suited for 3D samples, for which tuning the X-ray energy slightly below the absorption edge allows to sharply decrease absorption, while reasonably preserving magnetic scattering\cite{bib-NEE2024}. Other imaging techniques based on scattering exist, such as X-ray Fourier transform holography\cite{bib-EIS2004}, making use of the interference with a reference beam going through a hole or a slit, however they generally do not allow for sample scanning. Most of the above techniques, from (S)TXM to scattering, can be extended to tomography or similar approaches, making use of one or more series of sample images with varying angles to reconstruct volumic information\cite{bib-DON2017}. Overall, X-ray imaging techniques have a prominent position in advanced magnetic imaging techniques\cite{bib-CHR2024}.

Synchrotron radiation is naturally suited to implement time resolution of reproducible processes in a pump-probe scheme, as electrons come as bunches spaced typically by a few nanoseconds, producing periodic bursts of X-rays coming at the so-called cavity radio-frequency~(RF). Any measurement technique can harvest this underlying time structure by exciting the sample at the RF frequency or any of its harmonics. The time evolution of magnetization dynamics can be reconstructed by setting a controllable delay between pump and probe, over one period of excitation. For magnetic imaging this has been pioneered using photo-emission electron microscopy\cite{bib-VOG2003}, and since then applied to many other techniques\cite{bib-PUZ2010,bib-BUS2010,bib-FIN2018}, including a few with ptychography\cite{bib-BUT2025b} and also combined with 3D vector-field reconstruction\cite{bib-DON2020}. It is this protocol that we have implemented on the Ptychography setup of the HERMES beamline at SOLEIL synchrotron, and reported in the present manuscript to demonstrate its versatility in terms of both dynamics phenomena and range of time scales and frequencies. Note that, as it is based on a 2D detector, ptychography cannot be combined with register-based fast-counting electronics to implement time resolution with arbitrary periods of patterns, in contrast to STXM such as implemented at SLS and BESSY synchrotrons\cite{bib-FIN2018}.

\section{Methods}
\subsection{Ptychography at the HERMES beamline}

Time resolved ptychographic measurements were performed at the scanning transmission X-ray microscopy (STXM) end-station of the HERMES beamline at SOLEIL synchrotron. HERMES is a soft X-ray beamline designed for spectromicroscopy (STXM and XPEEM), operating over a large photon-energy range from \qtyrange{70}{2500}{\electronvolt}. Two APPLE-II undulators, HU64 and HU42, allow to cover the energy range with full control of the light polarization. This is particularly useful for magnetic studies, since measurements can be performed with opposite circular polarizations to access X-ray magnetic circular dichroism (XMCD), or with orthogonal linear polarizations to investigate X-ray linear dichroism (XLD). In the STXM branch, the beam is focused onto a set of slits that act as the secondary source for the focusing optics of the microscope. A more detailed description of the beamline design and optical characteristics can be found elsewhere\cite{bib-BEL2015}.

The STXM microscope is based on Fresnel zone plate optics, which focus the monochromatic X-ray beam onto the sample. The specimen is mounted on an XYZ piezoelectric stage, allowing precise positioning and scanning with respect to the focal spot. The relative position of the sample and zone plate is stabilized by a differential laser interferometer, providing a positional stability better than \qty{2}{\nano\meter}. For the ptychography measurements, a Fresnel zone plate with outermost zone widths of \qty{50}{\nano\meter} has been used. For ptychographic acquisition, a two-dimensional soft X-ray-sensitive camera mounted on the detector stage is used. The camera used in the HERMES implementation is a customized, vacuum-compatible Tucsen Dhyana95 CMOS system equipped with a backside-illuminated GPixel sensor\cite{bib-DES2020c}. The detector has a $2048 \times 2048$ pixel array, with a pixel size of $11 \times \qty{11}{\micro\meter\squared}$. The detector position can be adjusted to center the diffraction pattern and to select the sample-to-detector distance, which determines the reciprocal-space sampling and the maximum scattering angle collected. Further details of the detector design and its integration into the HERMES STXM microscope are described elsewhere\cite{bib-DES2020c}. During the acquisition procedure and at each sample position, a two-dimensional far-field diffraction pattern is recorded. The typical scan size is $1.5\times\qty{1.5}{\micro\meter\squared}$. The distance between consecutive positions is chosen to provide sufficient X-ray beam overlap (30-50$\%$) between neighboring illuminated areas, which is essential for a reliable ptychographic reconstruction.

The diffraction data are further reconstructed using PyNX, an open-source Python toolkit developed for coherent X-ray imaging and providing GPU-accelerated algorithms for ptychography\cite{bib-FAV2020}. Before reconstruction, the detector background is measured with the X-ray beam switched off and further subtracted from the diffraction patterns. PyNX uses the measured diffraction intensities together with the scanning geometry to iteratively recover the complex transmission function of the specimen and the illuminating probe. The final complex reconstruction provides complementary amplitude and phase images. For measurements performed at different photon energies or polarizations, the reconstructed images are aligned to correct any experimental drift. The amplitude is converted into optical density, and the resulting images can be combined to obtain spatially-resolved magnetic dichroic contrast image~(XMCD maps). All measurements in the present manuscript have been recorded with circularly left/right polarized light. The energy was set about \qty{1}{\electronvolt} below the Fe L$_{3}$ (\qty{707}{\electronvolt}) edge, as a balance between the dichroic response and transmission\cite{bib-NEE2024}.

\subsection{Samples}

We have chosen a type of sample that enabled us to evidence a number of different physical phenomena: DW motion, emission of spin waves, susceptibility of domains, and switching of magnetization. These consist of magnetic nanowires, whose stimulus is the \OErsted field arising from an electric current fed directly in the wire. While this situation is specific, the phenomena described in the present demonstration could be sought in other situations, with stimulus from spin-transfer torque, strip lines or others. More specifically, we considered permalloy \ce{Fe_20Ni_80} cylindrical nanowires ~(NWs) with composition modulations \ce{Fe_70Ni_30} localized along their length\cite{bib-RUI2018}, synthesized via template-assisted electrochemical deposition under conditions described in Ref.\cite{bib-FRU2026b}. After growth, the NWs were released by chemical dissolution of the alumina template, dispersed onto Si substrates with \qty{20}{\nano\meter}-thick  $100\times\qty{100}{\micro\meter}$ \ce{Si_3N_4} windows to enable X-ray microscopy experiments. To allow charge current to flow through the NWs, we contacted them using laser lithography with two electrodes separated by typically \qtyrange{20}{25}{\micro\meter}. The measured resistivity was approximately $\qty{15}{\micro\ohm\centi\meter}$. In all cases, after electrical contacting, the devices were covered with \qty{20}{\nano\meter} of alumina by atomic layer deposition to improve heat transfer from the wire to the supporting nitride membrane.

\begin{figure}[h]
  \centering
  \includegraphics[width=88mm]{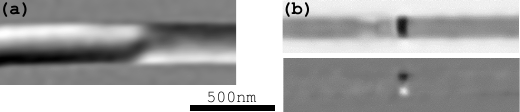}
  \caption{Structural and magnetic ptychography of permalloy wires at rest: (a)~First series, with diameter \qty{245}{\nano\meter} and modulation length \qty{60}{\nano\meter}~(no modulation shown on this image) (b)~Second series, with diameter \qty{135}{\nano\meter} and modulation length~\qty{100}{\nano\meter}, with transmission~(top) and XMCD~(bottom) images\dataref{2020-12-16-86-87}.}
  \label{fig:samples}
\end{figure}

A first series of NWs has a diameter of \qty{245}{\nano\meter}, without or with chemical modulation. In the latter case, the length of the permalloy segments is \qty{3.6}{\micro\meter}, and the length of modulations is \qty{60}{\nano\meter}. For this diameter, the wires display a vortex state, consisting of longitudinal magnetization on the axis and an azimuthal tilt of magnetization at its periphery\cite{bib-GOM2026}. The magnitude of curling at the periphery depends on the growth conditions, annealing with current-pulses or capping with atomic layer deposition\cite{bib-FRU2027b}. The samples are first magnetized longitudinally, so that magnetization on the axis is uniform along the entire wire. Domain walls~(DWs) may arise at the periphery of the wire, separating two areas with opposite circulation\cite{bib-FRU2026b}\bracketsubfigref{fig:samples}a. If present, the chemical modulations act as preferential pinning site for these DWs.

A second series of samples has a diameter of \qty{135}{\nano\meter}, permalloy segment length \qty{4.5}{\micro\meter} and modulation length~\qty{100}{\nano\meter}\dataref{Fairey F7, experiments ptycho SOLEIL 2025-10}. For this diameter, the segments are uniformly magnetized along the wire axis. These samples are also first magnetized along their axis, so that they do not display DWs. The chemical modulations give rise to a curling of magnetization, that allow to reduce the magnetostatic energy resulting from the magnetization mismatch between the two materials\cite{bib-FRU2022}\bracketsubfigref{fig:samples}b.

In both cases the collection of wires in solution involves permanent magnets whose stray field exceed by far the coercive field of the nanowires. Consequently, in the as-prepared state the component of longitudinal magnetization of wires of both series~1~(magnetization on the axis) and~2~(the entire magnetization) is uniform, with no DW.

\subsection{Electronics}

We excite the samples with a charge current fed directly in the wires. This current gives rise to an \OErsted field, which  couples directly with the azimuthal magnetization texture of the sample. For the samples of the first series, the \OErsted field  can induce the motion of DWs, the nucleation of a reversed domain or even the full azimuthal switching via coherent rotation\cite{bib-FRU2026b}. For samples of the second series, the \OErsted field can switch the circulation of the modulations\cite{bib-FRU2022}.

In the time-resolved pump-probe measurements, the current fed into the wire serves as the pump. The sample is mounted and wire-bonded on a dedicated PCB equipped with SMP connectors, with cables ending with SMA connectors fitted in a port of the STXM vacuum chamber. We used two instruments  to feed and control current into the wire. The first instrument is a two-channel arbitrary wave generator~(AWG), model M8195 from Keysight. It has a maximum output voltage of \qty{500}{mV}, and sampling rate \qty{65}{GSa/s}. The second instrument is a two-channel pulse generator, model PG1502 from Active Technologies. It has a rise and fall time slightly below a nanosecond, has a maximum output of $\qty{50}{\volt}$, with a triggering frequency up to \qty{40}{\mega\hertz}. The input and output signals were monitored using a four-channel oscilloscope, model UXR from Keysight. Its bandwidth is \qty{10}{\giga\hertz} and sampling rate \qty{65}{GSa/s}.

\begin{figure}[h]
    \centering
    %\fbox{\rule{0pt}{5cm} \rule{8cm}{0pt}}
    \includegraphics[width=83.644mm]{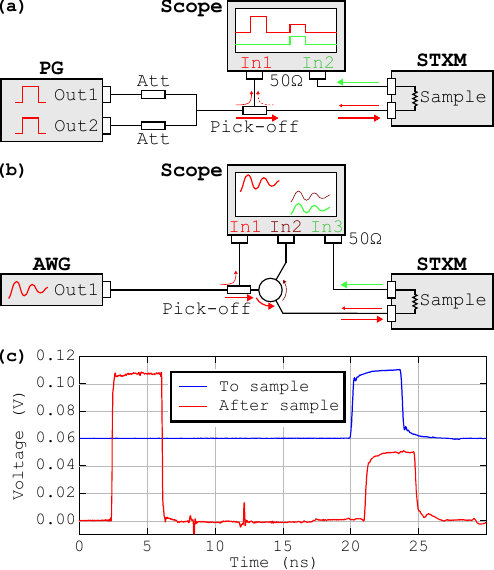}
    \caption{(a)~Electric circuit for the use of a two-channel pulse generator~(PG, or alternatively an arbitrary wave generator), with attenuators, a tee or resistive power combiner, a pick-off tee, the sample and an oscilloscope. (b)~Electric circuit for the use of a continuous exciting waveform, including an additional circulator (c)~Sent, reflected and transmitted signal for a square pulse of width $\qty{3.5}{\nano\second}$ sent to a sample with resistance \qty{5}{\ohm}. The two curves are offset for clarity. The open port of the protective tee is responsible for the derivative-like features seen on the first channel.}
    \label{fig:electric-circuit}
\end{figure}

\subfigref{fig:electric-circuit}a shows the wiring setup in case a pulse generator is used. The two channels of the pulse generator can be combined, \eg, a positive and negative pulse with a delay to excite and reset the sample, respectively. The attenuators are inserted before the resistive power combiner~(or tee) to lower the power returning to the pulse generator via the other channel. On the way to the sample, a small fraction of the power is fed to the oscilloscope with a resistive pick-off tee to monitor the magnitude of the voltage pulse sent to, and reflected from, the sample. Finally, the current at the output of the sample is fed into the oscilloscope. In the case of the continuous feeding of a sinusoidal waveform, \eg, through an AWG, a circulator is inserted between the pick-off tee and the sample\bracketsubfigref{fig:electric-circuit}b, to avoid the occurrence of a standing wave due to the reflected power arising from the mismatch of the sample resistance. In both cases, we often insert a tee connector between the pick-off tee and the entrance to the chamber. The cables from the generator and to the sample fit two port. A short can be fit to the third port to protect the sample against electrostatic discharged when handling the pulse generator, the AWG or the circulator. This third port is left open during the experiments, with only a weak parasitic effect\bracketsubfigref{fig:electric-circuit}c.

Prior to exciting the sample and conducting pump-probe measurements, we measure the magnitude of the incoming, transmitted and reflected voltage of isolated nanosecond-long square pulses sent to the wire. This directly reflects the current flowing in the wire, related to the \qty{50}{\ohm} impedance of the cable. From the missing voltage~(reflected and transmitted versus incoming) we evaluate the losses, which are in general a few percent at most. From the height of the plateaus we derive the wire resistance, and check that it matches the one measured in dc prior to its mounting:
\begin{equation}
    V_\mathrm{out}=R_\mathrm{out}I=\frac{2R_\mathrm{out}}{R_\mathrm{gen}+R_\mathrm{s}+R_\mathrm{out}}V_\mathrm{gen}\;,
    \label{eq:wire-resistance}
\end{equation}
with $R_\mathrm{gen}=\qty{50}{\ohm}$ the generator internal resistance, $R_\mathrm{out}=\qty{50}{\ohm}$ the cable impedance/oscilloscope input impedance after the sample, $R_\mathrm{s}$ the sample resistance and $V_\mathrm{gen}$ the voltage request to the generator. The voltage of the transmitted signal, converted through the \qty{50}{\ohm} input of the oscilloscope, is used as a  probe of the current flowing in the wire during an experiment. This also allows to evaluate the broadening and distortion of the rise and fall fronts. The decrease of transmitted~(or increase of reflected) current allows to estimate the rise of temperature of the sample, based on the temperature-dependent resistance of the wire\dataref{Pulse-generator-2025-10-07.pptx}. This applies to the gradual increase of temperature during a pulse, or the average and constant rise for an ac excitation. In the frequency space, we evaluate the transmitted current via the magnitude of a sine wave generated by the AWG, versus its frequency.

The RF frequency of the storage ring at SOLEIL is $f_\mathrm{RF}=\qty{352.202}{\mega\hertz}$. Therefore, in the full filling mode the bunches of X-rays arrive every other $T_\mathrm{RF}=\qty{2.839}{\nano\second}$. A TTL-like signal from the ring is provided to the user via the so-called Timbel interface card on the beamline, which can be set to a harmonics of the RF frequency, up to the ring frequency. The delay of this reference signal with respect to the ring can be adjusted down to steps of \qty{10}{\pico\second}. The reference signal is fed into the users' electronics, to serve either as a trigger for the pulse generator, or as an external clock for the AWG and lock the phase of its output signal. The full width at half maximum of the X-ray bunches is \qty{47}{\pico\second}~(\qty{18}{\pico\second} root-mean-square, RMS) in the full filling mode. The pulse-to-pulse arrival-time jitter is on the order of a few ps RMS, so the effective time resolution is essentially set by the pulse width itself, in total on the order of \qty{50}{\pico\second}. SOLEIL also occasionally provides 8-bunch or 1-bunch fillings modes, with specifications gathered in \tabref{tab:fillingModes}. The full-filling mode allows to maximize the flux available for imaging, which is not different from under static conditions. However, periodic excitations such as sine waves are restricted to harmonics of the RF frequency, which may not provide sufficient sampling~(in the frequency space) or duration~(in the time space) for many phenomena of magnetization dynamics. Low-filling modes circumvent this limitation at the expense of lower flux. These also allow the sample to cool down between two excitations.

\begin{table}[h]
    \caption{Specifications of the various filling modes at SOLEIL}
    \label{tab:fillingModes}
    \begin{tabular}{lcccccc}
      \hline\hline
      % after \\: \hline or \cline{col1-col2} \cline{col3-col4} ...
       & Bunches & $f$ (\si{\mega\hertz}) & $T$ (\si{\nano\second}) & Bunch FWHM (\si{\pico\second}) &  $I$ (\si{\milli\ampere} \\
      \hline
      Full filling & 416 & 252.202 & 2.839 & 47 & 500 \\
      Eight bunch & 8 & 6.7731 & 147.64 & 89 & 100 \\
      Single bunch & 1 & 0.8476 & 1179.8 & 98 & 20 \\
      \hline\hline
      \hline
    \end{tabular}
\end{table}

\section{Results}

\subsection{Foreword}

The purpose of the following paragraphs is to describe the methods and illustrate the potential of time-resolved ptychography to address timely aspects of nanomagnetism and spintronics. While we introduce the context of the underlying physical questions, and show a result gained thanks to time-resolved ptychography, it is not the purpose of the present manuscript to analyze in-depth the various phenomena taking place. Physical results will be reported in more detail separately.

We have chosen a single sample platform to commission time resolution at the HERMES beamline: single cylindrical magnetic nanowires with diameter around \qtyrange{130}{250}{\nano\meter}. From a scientific perspective, the inherent limitations of 2D logic and storage density have indeed led to a growing interest in devices exploiting the third dimension. As regards data storage, a fascinating proposal is the so-called race-track memory\cite{bib-PAR2004,bib-PAR2008}, whose concept is to code information in long and vertical magnetic nanowires in the form of magnetic DWs. From a fabrication point of view, such architectures can be realized via deep physical or chemical etching and filling, naturally delivers cylindrical structures, \eg, wires or tubes\cite{bib-SOU2014,bib-FRU2018d,bib-VAZ2020}. Such systems fall in the broader field of 3D nanomagnetism\cite{bib-FRU2017b}, an emerging research area in which curvature and topology influence static and dynamics magnetic properties. Experimental realizations in 3D are emerging, and ptychography seems  well positioned to contribute to their characterization, as these objects are very efficient scatterers.

\subsection{Domain-wall motion}
\label{sec:dw-motion}

Magnetic DWs have emerged as promising information carriers in both logic\cite{bib-LUO2020} and memory\cite{bib-PAR2015b} devices. The understanding of the physics of DW motion and its control are pivotal for the development of devices. Starting more than two decades ago, theory and simulation predicted a number of physical effects specific to DWs in cylindrical systems, such as DWs embedding the Bloch point micromagnetic singularity at rest\cite{bib-FOR2002,bib-HER2002a}, its much-delayed Walker breakdown upon motion\cite{bib-THI2006}, and the possibility to reach the magnonic limit\cite{bib-YAN2011b,bib-HER2016}, for which the DW gets dressed by spin waves. Experiments have confirmed the existence of Bloch-point DWs\cite{bib-FRU2014}, their energy landscape in curved situations\cite{bib-RUI2025}, their motion under quasistatic magnetic field\cite{bib-FRU2019}, their potential for reaching speed of several hundreds of \si{\meter/\second} when driven by spin-transfer torque, along with the key role played by the \OErsted field\cite{bib-FRU2019b}, the possible change of DW topology upon motion\cite{bib-FRU2019,bib-FRU2025}. In these examples DWs are monitored under static conditions by magnetic microscopy, the DW speed being inferred from the ratio of the distance between its position before and after application of a pulse of magnetic field or electric current, over the length of the pulse\cite{bib-FRU2019b,bib-BRA2023}. This approach is easy to implement, but may suffer from biases if DW motion is not steady-state during the pulse, because of inertia, pinning, transformations etc. Time-resolved imaging opens the possibility to directly track DW motion, unraveling the intimate features of the dynamics and providing a more reliable determination of DW speed. However, such measurements introduce the constraint that the underlying dynamics processes need to be reproducible billions of times.

\begin{figure}[h]
    \centering
    %\fbox{\rule{0pt}{5cm} \rule{8cm}{0pt}}
    \includegraphics[width=83.663mm]{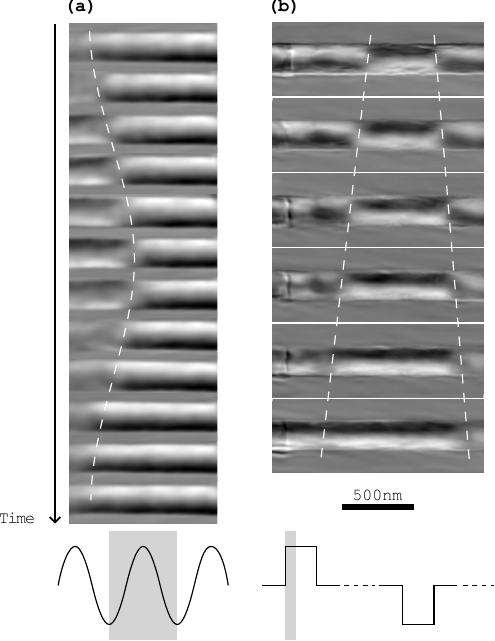}
    \caption{Series of ptychography XMCD images of a nanowire with composition \ce{Fe_70Ni_30} and diameter \qty{245}{\nano\meter}. (a)~Images taken in full-filling mode under sinusoidal excitation. The delay between images is $\qty{236}{\pico\second}$. $J = \qty{4E10}{\ampere\per\meter\squared}$ and $f = \qty{352.202}{\mega\hertz}$ (radio-frequency)\dataref{Velvet V8 2024-10-06-109to134} (b) Images taken in 8-bunch mode under pulsed excitations with $J = \qty{7E11}{\ampere\per\meter\squared}$ and $t=\qty{5}{\nano\second}$. The delay between two images is $\qty{250}{\pico\second}$. This wire was capped with \qty{20}{\nano\meter} \ce{Al_2O_3} ALD and the backside of the Si nitride window with \qty{100}{\nano\meter} Al.\dataref{Bruital B1,  2025-10-10-141-to-154} In both cases the XMCD contrast is $\pm\qty{15}{\%}$, and dotted white lines are guide to the eye  for DW location. The waveform of the injected current is schematized at the bottom, with the gray area highlighting the span of the images. Time flows from the top to the bottom row.}
    \label{fig:DWs}
\end{figure}

Here, we considered wires from series~1, displaying azimuthal magnetization. First, we send a single current pulse, with typical magnitude \qty{5E11}{\ampere\per\square\meter} and duration of a few nanoseconds. This sets a given and single circulation of azimuthal magnetization in the entire wire. Second, a series of single nanosecond current pulses with opposite sign and raising magnitude is sent until one or more domains with opposite circulation are nucleated. Starting from this initial state, \subfigref{fig:DWs}{a} shows a series of XMCD images taken in full-filling mode at different delays, while continuously exciting the sample with a sinusoidal current of amplitude $J = \qty{4E10}{\ampere\per\meter\squared}$ at the RF frequency $f = \qty{352.202}{\mega\hertz}$. These images illustrate magnetization dynamics over one full period of the excitation. As seen in the images, the DW moves following the oscillatory excitation, consistent with the effect of the \OErsted field. From this series, the maximum velocity is determined to be $v =\qty{640}{\meter\per\second}$, reached during around \qty{500}{\pico\second} when the DW is in its central position. Following the motion of a DW in steady-state and over longer times requires to work in a partial-filling mode. \figref{fig:DWs}(b) illustrates DW motion tracked with an 8-bunch mode, with square pulses of current with amplitude $J = \qty{7E11}{\ampere\per\meter\squared}$ and duration $t=\qty{5}{\nano\second}$. In this case, the delay between two consecutive images is $\qty{250}{\pico\second}$. The full sequence of current consists of two pulses of opposite sign to move the DW back and forth in a reproducible fashion, separated by a rest time to allow the sample to cool down. The two dashed lines are guides to the eye to indicate the position of the two DWs over time. Consistent with the expected effect of the \OErsted field, two consecutive DWs move along opposite directions. We believe that the unusually-high speed for field-driven DW motion under these two protocols arises from an exchange-spring effect between the DW at the wire periphery, and magnetization robustly lying along the longitudinal direction on the wire axis. A more detailed analysis is provided elsewhere\cite{bib-FRU2026b}. Note that in both series some areas of the wire appear homogeneously gray during the magnetization process, \eg, the first two frames in \subfigref{fig:DWs}{a}. This may result from either a cancellation of azimuthal magnetization as a balance between azimuthal anisotropy and the \OErsted field, or the averaged superposition of different states during the dynamics, resulting from the non-reproducibility of the magnetization process.

The high spatial resolution of ptychography may help identifying fine changes in the internal structure of the DW during its dynamics, such as distortion and change of width. This should allow to go beyond the usual approach of analyzing the DW speed only.

\subsection{Fast magnetization switching}
\label{sec:switching}

Curling is a common way for soft-magnetic materials to close their flux and thereby reduce their magnetostatic energy. The concept had been recognized for long in the context of the Brown paradox\cite{bib-FRE1957,bib-BRO1958}. In flat elements, an example of broad significance is the magnetic vortex\cite{bib-MER1979,bib-COW1999b,bib-SHI2000b}. In 3D systems the incentive to close the magnetic flux is even larger, owing to the stronger and long-range magnetostatic interactions. The transition to the vortex state occurs at a few times the dipolar exchange length\cite{bib-SCH1988}, resulting from the competition between the gain in magnetostatic energy and the cost in exchange. In cylindrical nanowires with longitudinal magnetization, this may take the form of so-called end curling states at both apex\cite{bib-ZEN2000,bib-WAN2008a}, longitudinal curling at material modulations\cite{bib-FRU2022}, longitudinal and combined longitudinal/transverse curling in Bloch-point and Transverse-Vortex DWs, respectively\cite{bib-FRU2015b}. The curling structures exist in both their left- and right-circulation forms, which are degenerate at rest unless coupled by exchange with another nearby curling structure\cite{bib-WYS2017}. A pulse of charge current fed into the wire lifts the degeneracy between two longitudinal states with opposite circulation, related to the sign or the associated \OErsted field. It has been reported that above a given threshold of current density the circulation can be switched, \eg, for a Bloch-point wall\cite{bib-FRU2019b} or a chemical modulation\cite{bib-FRU2022}. Although switching has been reported for current pulses down to the nanosecond time scale, their real-time dynamics has not been reported so far.

\begin{figure}
    \centering
    \includegraphics[width=71.531mm]{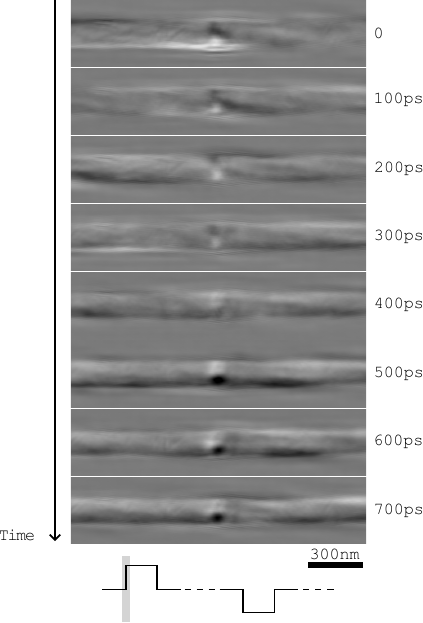}
    \caption{Series of ptychography XMCD images of a nanowire with composition \ce{Fe_20Ni_80}, diameter \qty{135}{\nano\meter} and \qty{100}{\nano\meter}-long chemical modulations of composition \ce{Fe_70Ni_30}~(series~2) during the application of a current pulse with density \qty{3.6E11}{\ampere\per\square\meter} and duration \qty{3}{\nano\second}. The waveform of the injected current is schematized at the bottom, with the gray area highlighting the span of the images. Time flows from the top to the bottom row. Two consecutive images are separated by \qty{100}{\pico\second}. All images are displayed with the same XMCD contrast of \qty{15}{\%}.\dataref{Fairey-F7-2025-10-12-072-096}}
    \label{fig:switching}
\end{figure}

We have considered samples of series~2, \ie, with composition \ce{Fe_20Ni_80} in the segments, \qty{100}{\nano\meter}-long chemical modulations of composition \ce{Fe_70Ni_30}, and diameter \qty{135}{\nano\meter}. Using static imaging before and after sending pulses of current with nanosecond duration \cite{bib-FRU2022}, we have determined that the threshold for switching circulation in  large-magnetization modulations of these particular wires is~\qty{3E11}{\ampere\per\square\meter}. The pulse sequence implemented to image switching in real time consists of two pulses of same duration \qty{3}{\nano\second} but opposite sign, so that switching occurs twice during every period. The two pulses are separated by a rest time, to allow the wires to cool down. Consequently, this has been implemented under the 8-bunch filling mode of the ring. \figref{fig:switching} shows a series of ptychography XMCD images, each separated by \qty{100}{\pico\second}, during one of the two pulses of magnitude~\qty{3.6E11}{\ampere\per\square\meter} and duration~\qty{3}{\nano\second}. This series shows that switching of circulation at the modulation is achieved in a few hundreds of picoseconds (\figref{fig:switching}). The physics underlying this very fast field-driven switching will be detailed elsewhere\cite{bib-FRU2027c}.

Ptychography allows us to resolve not only the occurrence of switching, but also details about the inner structure of the magnetization texture during the process, thanks to its high spatial resolution. For instance, the width~(measured along the wire axis) and magnitude of the contrast appear smaller during \qty{200}{\pico\second}, providing hints about the detailed switching mechanism. Besides, for such processes where a waiting time of ten(s) of nanoseconds is required for sample cooling, working with an 8-bunch mode is more advantageous than with a full-filling mode such as implemented for STXM\cite{bib-FIN2018}. Indeed, although the total beam flux is reduced fivefold, in the 8-bunch mode we can focus on the process of interest lasting less than \qty{1}{\nano\second} out of the $\sim\qty{150}{\nano\second}$, while in the latter case the process is evenly sampled over the period of the excitation, during which the sample is a rest most of the time. In  this particular case,  the 8-bunch mode provides a $30:1$ advantage. This can be used to gather larger images, or gain in the signal-over-noise ratio.

\subsection{Spin waves}
\label{sec:sws}

Non-reciprocal propagation of spin waves~(SWs) in 3D nanosystems due to curvature has been predicted, \ie, with a $\omega(k)$ dispersion curve that is  non-symmetric for opposite $\vect k$ vectors\cite{bib-OTA2016}. The seminal prediction in Ref.\cite{bib-OTA2016} concerns nanotubes and considers propagation along the tube axis. In short, the physics is the following: when magnetization is azimuthal the SW has the so-called Damon-Eshbach geometry, \ie, magnetization and the wave vector~$\vect k$ (here along the tube axis) are perpendicular. For thin films with this geometry, it is well known that spin waves tend to be localized at surfaces, where volume magnetostatic charges can be largely screened by surface charges, while the SWs are evanescent in the volume of the film\cite{bib-ESH1960,bib-DAM1961,bib-STA2009}. SWs propagating along $+\vect k$ or $-\vect k$ are localized at opposite surfaces of the film, which results from the phase of volume charges being $\pi$-shifted for opposite~$\vect k$, matching the surface charges that are opposite on the two opposite sides of the thin film. If both film surfaces are equivalent, then the dispersion curve is symmetric. However, if opposite surfaces experience different conditions, \eg with a Dzyaloshinskii-Moriya interaction or in a synthetic antiferromagnet\cite{bib-MIL2023}, then the dispersion curve is non-symmetric. Curvature in a 3D system, such as the rolled sheet making a nanotube, also lifts the degeneracy for opposite $\vect k$ vectors. The first reason is that the lengths of the inner and outer surfaces are different, with possible impact on both the densities of exchange and magnetostatic energies. The second reason is that each surface faces a different environment, \ie for a nanotubes, the inner one facing the other parts of the tube, and the outer one facing free space, with an impact on the strength of magnetostatic interactions.

Only a few experimental reports have followed these predictions, all conducted on nanotubes with a hexagonal cross-section and magnetization at least partially azimuthal. Investigations have been conducted both in real space using time-resolved STXM\cite{bib-KOE2021}, and in the frequency space using micro-focused Brillouin Light Scattering\cite{bib-GIO2023}. Both provide indications of the occurrence of non-reciprocity, inferred directly from magnetic images for the former, and indirectly via micromagnetic simulations for the second. However, the analysis of the experimental data is made difficult, because of the complexity of the system: curvature is not uniform in the tube, consisting of flat facets, separated by edges with a strong curvature. Therefore, examining spin waves on rotationally-invariant systems is desirable.

Here, we consider again wires of series~1, \ie, displaying magnetization largely azimuthal at the wire periphery. The protocol to nucleate DWs is similar to the one described in \secref{fig:DWs}. Then, we feed a high-frequency ac current in the wire, with purpose to set the DWs in an oscillatory motion, under the stimulus of the high-frequency \OErsted field. We expect that this oscillatory motion can act as a source of spin waves, in a fashion similar to a magnetic vortex\cite{bib-WIN2016} or DWs\cite{bib-MAY2021,bib-HOL2018} in thin films or flat elements. In practice, the frequency of the current is set to the RF frequency $f_\mathrm{RF}=\qty{352.202}{\mega\hertz}$ or one of its harmonics, with phase locked to the one of the synchrotron ring. We applied a current magnitude up to \qty{7E10}{\ampere\per\square\meter} and with frequency from \qty{352}{\mega\hertz} up to \qty{4.93}{\giga\hertz}, although results are shown here for only one value of each.

\begin{figure}[h]
    \centering
    %\fbox{\rule{0pt}{5cm} \rule{8cm}{0pt}}
    \includegraphics[width=135.744mm]{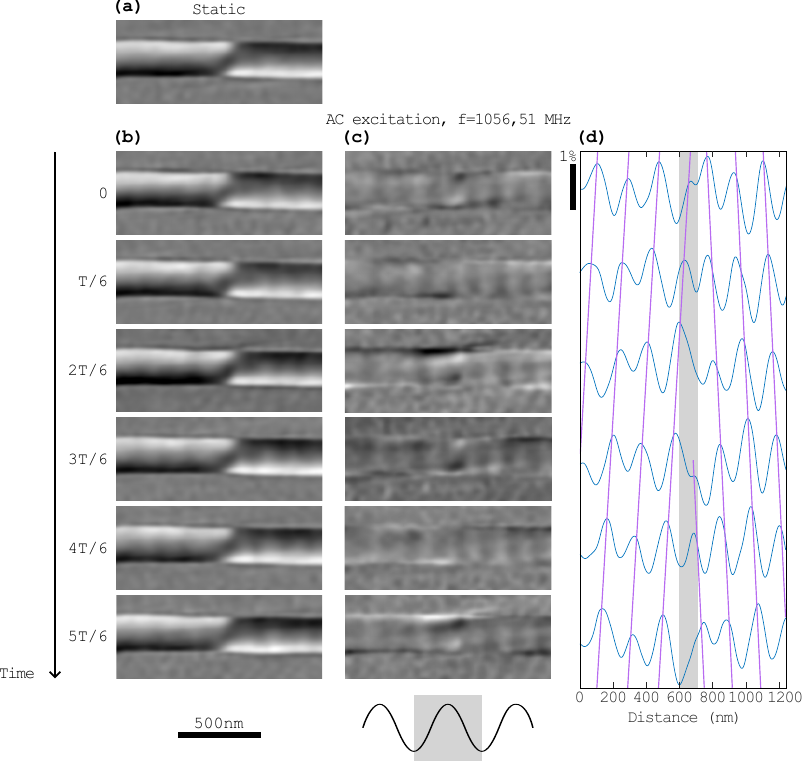}
    \caption{(a)~Static XMCD image of a nanowire with composition \ce{Fe_20Ni_80}, diameter \qty{245}{\nano\meter}~(series~1),  displaying a vortex state and a DW in the field of view. Starting from this state, an ac current with magnitude \qty{5.5E10}{\ampere\per\square\meter} and frequency \qty{1056.61}{\mega\hertz} is continuously fed in the wire, and imaging is conducted in the full-filling mode of the synchrotron ring. The lapse between two frames is one sixth of the RF period, time flowing from the top to the bottom image. (b)~Ptychography XMCD images, computed as the normalized difference of two images with opposite X-ray polarization. The range of contrast is $\pm0.06$ (c)~Ptychography XMCD images of the periodic dynamic component magnetization, computed as the normalized difference of two images with the same X-ray polarization by $\pi$-shifted in time. The range of contrast is $\pm0.03$ (d)~Longitudinal cross-sections of the same, averaged over a width of $\qty{34}{\nano\meter}$. The lines are guides to the eyes to identify the maxima in the cross-sections.\dataref{2024-12-07-073-084}}
    \label{fig:SWs}
\end{figure}

\subfigref{fig:SWs}b shows a time-resolved series of ptychography XMCD images illustrating the emission of spin waves from an oscillating DW, at the third harmonic of the RF frequency~(\qty{1056.61}{\mega\hertz}). While oscillatory patterns are visible in these images, they can be made more visible by computing the difference of two images taken with the same circular polarization of X-rays but shifted in time by half a period\bracketsubfigref{fig:SWs}c, which highlights the periodic part of dynamic magnetization. Longitudinal cross-sections of the latter\bracketsubfigref{fig:SWs}{d} clearly show that the wave moves away from the DW by one wavelength \qty{170}{\nano\meter} over one period, as expected for the emission of spin waves. SWs are seen on either sides of the DW, which can be understood as the Damon-Eshbach conditions are the same: both $k$ and the circulation of azimuthal magnetization are switched to their opposite. When moving to the next DW to either the left or the right~(say, labeled $n-1$ and $n+1$), we have checked that SWs seem to move towards the DW. However, backward propagation is not possible as what is observed with an ac excitation is the phase velocity, which is always positive. We believe that we are instead imaging the SWs arising from the neighboring DWs~(say, labeled $n-2$, $n$ and $n+2$), outside the field of view, and that no SW is emitted from one of two DWs. We have confirmed with micromagnetic simulations using the TetraX software\cite{bib-KOE2021a} that a band gap is indeed open for one direction of propagation for a given set of magnetization circulation around the DW~(say, left-right or right-left). This contrasts with the weaker non-reciprocal effect predicted for tube, inducing only a distortion of the dispersion curve, which can be related to the large amount of magnetic material at the core of the wire, compared to a tube. A more detailed analysis will be provided elsewhere\cite{bib-FRU2027}.

The high spatial resolution of ptychography provides several decisive advantages. First, it should allow to detect spin waves with wavelength down to \qty{25}{\nano\meter} or so, which is comparable to that of SWs created by moving DWs in tubes in the so-called magnonic regime\cite{bib-YAN2011b}. Second, even for slightly larger wavelength, it limits the averaging of contrast along the propagation direction, which allows for quantitative analysis of the SW amplitude. Third, the fine structure of the DW may be monitored during the oscillation, and allow for finer modeling of the emission.

\subsection{Domain response, towards ferromagnetic resonance}
\label{sec-domainResponse}

In the previous section we have reported the imaging of spin waves, which are the non-zero-wave-vector form of ferromagnetic resonance. Beyond the situation of their excitation by a local source, like previously discussed, spin waves may exist in extended systems, either thermally activated, or uniformly excited such as in ferromagnetic resonance. Spin waves in cylindrical nanowires have been the focus of theoretical studies over the past 25 years\cite{bib-ARI2001,bib-ARI2004,bib-RYC2018}, pointing at modes quantized both along the wires radius and its perimeter, therefore labeled with two indexes. A number of experimental works followed soon\cite{bib-EBE2001b,bib-WAN2002}, however their analysis could not be quantitative, because the measurements had been conducted on dense assemblies of nanowires strongly interacting via their dipolar fields. In a recent experiment we demonstrated the spectroscopic measurement of a half-dozen of the lowest-energy modes of a single and longitudinally-magnetized nanowire, using micro-focused Brillouin Light Scattering\cite{bib-FRU2025c}. Here, we illustrate the potential of time-resolved XMCD imaging to investigate such modes in real space.

\begin{figure}
    \centering
    \includegraphics[width=8.7313cm]{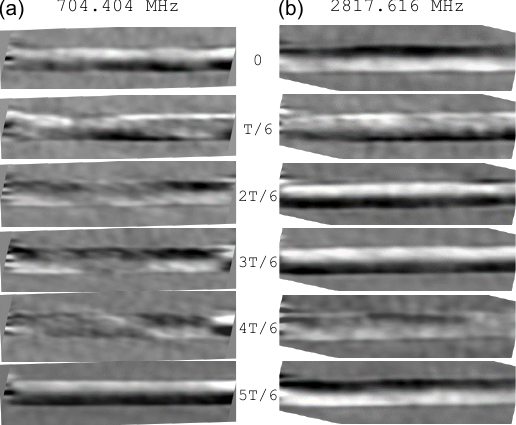}
    \caption{Time-resolved snapshots of the ac-\OErsted-field response of a permalloy nanowire with diameter \qty{135}{\nano\meter}, at frequencies (a)~$2f_\mathrm{RF}=\qty{704.404}{\mega\hertz}$ and (b)~$8f_\mathrm{RF}=\qty{2817.616}{\mega\hertz}$, for six different delays spanning evenly each period~$T$.}
    \label{fig:soleil-resonance}
\end{figure}

The lowest-energy mode in Ref.\cite{bib-FRU2025c} is the one labeled $(\ell=1,m=0)$ in terms of radial and azimuthal indexes. Note that in our notation, shared also by other authors in the field\cite{bib-RYC2018}, the azimuthal direction of magnetization is defined in the local frame $<\boldsymbol{\uprho},\boldsymbol{\upvarphi}>$, so that the index $m=0$ points at a rotationally-invariant mode. This mode couples naturally to the \OErsted field arising from an ac current fed into the wire as excitation. We considered wires of series~2, \ie, with composition \ce{Fe_20Ni_80} diameter \qty{135}{\nano\meter}, but no modulations\dataref{2024-12-SOLEIL-E7}. These display axial magnetization, and we checked in static conditions that magnetization is uniform in the field of view. We worked in the full-filling mode, feeding an ac current at a harmonic of $f_\mathrm{RF}$ with a density of $\qty{2.2e10}{\ampere\per\meter\squared}$, which translates in an exciting \OErsted field of $\qty{1}{\milli\tesla}$ at the periphery of the wire. The current decreased by about $\qty{30}{\%}$ at $\qty{2}{\giga\hertz}$, as measured from the transmitted signal, and could not be quantified at higher frequencies owing to the abrupt cutoff of the digital oscilloscope used at the time. Nevertheless, we investigated dynamically up to the eleventh harmonic of the cavity radio-frequency, \ie, from $\qty{352.202}{\mega\hertz}$ to $\qty{3885.222}{\mega\hertz}$. For each frequency we took six images equally spaced in time, in steps of $T/6$ with $T$ the excitation period. For each frequency the wire displays a black-white contrast across its width~\bracketfigref{fig:soleil-resonance}, reverting every half period or so, consistent with the expected response to the ac \OErsted field. However, we could not track the phase of the excitations in this experiment, so that we cannot discuss this response in relation to the resonance frequency, expected to be around \qty{3.7}{\giga\hertz} for this mode, based on micromagnetic simulations using TetraX. The actual resonant mode may however be lower, as we disregarded any contribution to azimuthal anisotropy, although at this diameter we are not far from the transition to azimuthal anisotropy. Besides, inhomogeneities along the wire axis were often visible, especially at low frequency. These may point at spatial inhomogeneities of anisotropy.

The application of time-resolved ptychography to ferromagnetic resonance get its limits with the intrinsic time resolution of the technique, limited to \qty{50}{\pico\second} at most, largely related to the FWHM of the electron bunch\brackettabref{tab:fillingModes}. Thus, frequencies larger than $\sim\qty{5}{\giga\hertz}$ will be difficult to investigate. The technique therefore rather suited to the resonance of magnetization textures, which is often lower in frequency, although in that case operation in the 8-bunch mode is required.

\section{Discussion and outlook}
\subsection{Use of the various ring filling}

The RF frequency of synchrotron facilities is in the sub-GHz regime: it is \qty{500}{\mega\hertz} at most rings, and \qty{352}{\mega\hertz} at SOLEIL. Thus, as it is not possible to gate cameras at this rate, full filling modes allow only to work with harmonics of this frequency, which are quite spaced in the frequency space. Nevertheless, this may be suitable to probe phenomena with a broad continuum, such as the dispersion curves of spin waves\bracketsecref{sec:sws}. The strength of ptychography is its good spatial resolution, enabling to resolve the somewhat large $k$ vectors at high frequency.

In contrast, low-filling modes offer a rest time of a few tens up to a hundred nanoseconds between two probes, or even around a microsecond for a single-bunch mode. This is crucial for many phenomena in nanomagnetism and spintronics. For instance, micromagnetic time scales often exceed the RF period, especially if a reset stimulus needs to be applied, as illustrated in \secref{sec:dw-motion} for DW motion. This is largely related to inertia, damping, and possibly dependent on the size of the sample in case of nucleation-propagation mechanisms\cite{bib-BAU2017}. Low-filling modes also open the possibility to monitor narrow resonances, providing a probing frequency comb with period a few MHz only. This applies to extended or localized resonances modes such as ferromagnetic resonance. Finally, the rest time may be crucial to let a physical device cool down in-between two excitations. This is particularly crucial for spintronics, for which the current flows directly into the physical system with current density up to \qty{1E12}{\ampere\per\square\meter}, such as illustrated in \secref{sec:switching}. We believe that time-resolved ptychography combined with low-filling modes will bring new insights, revealing fine details in the inner structure of dynamic magnetization structures or localized resonance modes, which may require accessing the length scale of the dipolar exchange length. Hybrid filling modes are not suitable for ptychography, due to the present inability to gate the pixels of cameras at the microsecond time scale.

The intermediate filling modes offer a good compromise between accessing the latter options, while maintaining a reasonable average flux. For instance, at SOLEIL the reduction of flux in the 8-bunch mode is only fivefold with respect to full feeling, meaning a reduction of S/N ratio of only slightly above 2. However, such modes are not or scarcely available in most synchrotrons. The spintronics community should support its broader availability.

\subsection{Time-resolved ptychography at 4th-generation facilities}

 Ptychography relies inherently on the transverse coherence of synchrotron radiation to extract phase information from the interference patterns of the scattered wavefield. While advanced reconstruction frameworks can accommodate a partial degree of coherence, it will obviously degrade fringe contrast and reduce the statistics of effectively coherent photons. The ongoing upgrade from third-generation synchrotrons to fourth-generation Diffraction-Limited Storage Rings (DLSRs) \cite{bib-BOR2014,bib-HET2014}, driven by multi-bend achromat (MBA) lattices, is poised to redefine the capabilities of coherent X-ray imaging. By drastically reducing the transverse emittance of the electron beam, MBA lattices boost the coherent fraction of the radiation, yielding a massive increase in the coherent photon flux reaching the sample. Projections suggest that MBA designs can deliver a two to three order of magnitude increase in average spectral brightness over legacy sources, even though the practical gain in coherent flux remains tightly bound to the photon energy, insertion device parameters, and beamline optics. Soft X-ray ptychography stands to highly benefit from this paradigm shift, as its quest for high spatial resolution hinges directly on photon statistics and spatial coherence\cite{bib-FIN2026}. Preliminary benchmarks at the HERMES beamline have already highlighted the power of this methodology for spectroscopic and chemical mapping, notably through carbon K-edge investigations and the characterization of nanostructured materials\cite{bib-BEL2015,bib-MIL2022}.

This influx of coherent photons opens new perspectives for stroboscopic time-resolved ptychography. At third-generation facilities, the broad horizontal emittance of the electron beam curtails the available coherent fraction in the soft X-ray regime, forcing researchers to accept long exposure times and extensive scan repetitions to accumulate enough photons at each pump-probe delay. These sustained measurements are highly vulnerable to mechanical drift, thermal fluctuations, and ambient instabilities. With the higher coherent flux of DLSRs, the exposure time per diffraction pattern can be compressed significantly. This speedup gives experimentalists the flexibility either to accelerate data collection or to enrich the dataset with finer spatial scanning grids and denser temporal delay points. Consequently, achieving nanometer-scale resolution becomes practically viable, smoother workflows follow, and tracking time-dependent magnetic textures becomes feasible. One must not, however, conflate a higher coherent flux with a guaranteed boost in spatial resolution. The final image quality remains bound by the maximum recorded scattering angle, detector noise, sample stability, and the robustness of the phase-retrieval algorithms. Similarly, an enhanced coherent fraction offers no suspension from radiation damage. The absorbed dose is still dictated by the cumulative total photon flux striking the sample, its intrinsic absorption cross-section, and the overall exposure conditions.

Beyond flux, the temporal landscape of the X-ray beam plays a decisive role. In standard storage rings, the duration of an X-ray pulse mirrors the longitudinal length of the electron bunch, which typically spans from \qtyrange{50}{100}{\pico\second} at third-generation sources, barring specialized short-pulse configurations\cite{bib-WEN2019,bib-HUA2014}. Because MBA lattices feature lower momentum-compaction factors, they can naturally support shorter electron bunches. The SOLEIL II reference design, for instance, targets a natural bunch length of roughly \qty{9}{\pico\second} RMS, translating to a Gaussian profile of about \qty{21}{\pico\second} FWHM under baseline operating conditions\cite{bib-GUB2024}. Such compressed pulses are ideal for capturing ultra-fast magnetization dynamics without losing structural details to temporal blurring during the X-ray probe windows. Still, a diffraction-limited lattice does not automatically guarantee shorter pulses. The real bunch duration is a complex function of RF voltage, momentum compaction, bunch charge, collective effects, and the potential deployment of harmonic cavities. The latter are frequently used to lengthen the bunch intentionally to safeguard beam lifetime and stability. Assessing the temporal resolution of a given DLSR therefore requires a close look at the specific filling pattern and operational mode chosen for the experiment. Maintaining an ultra-low emittance when the bunch charge is high introduces steep challenges in accelerator physics. Packing high charge densities into an electron bunch amplifies intra-beam scattering (IBS), an effect driven by multiple Coulomb collisions among electrons within the same bunch. IBS broadens both the transverse emittance and the energy spread, a degradation that reduces the coherent fraction of the photon beam and erodes the core advantage of a diffraction-limited infrastructure\cite{bib-KUB2005b}. Defining a viable operating envelope also means wrestling with other collective phenomena and Touschek scattering. These constraints are acutely felt in pump-probe experiments that rely on isolated, high-charge bunches to maximize the photon count per probe event, as these configurations tend to trigger unfavorable beam-dynamical responses. Optimizing the filling pattern, whether choosing single-bunch, few-bunch, or hybrid timing structures, involves balancing the number of photons per pulse, the repetition rate, the achievable bunch duration, and emittance preservation. The ongoing studies for SOLEIL II regarding harmonic-cavity performance and collective instabilities underline just how critical this balance is\cite{bib-GUB2024}.

Ultimately, the future of time-resolved ptychography cannot rely solely on accelerator upgrade.  The gain in coherent flux shifts the experimental bottleneck toward two-dimensional detector throughput, data acquisition, and computing power. Unlike standard scanning transmission X-ray microscopy, time-resolved ptychography demands saving a complete two-dimensional diffraction pattern at every single scan position and for each time delay. Multiplying this requirement across a series of pump-probe delays creates massive datasets that must be streamed, processed, and inverted without loss. Next-generation detectors must pair high frame rates with high dynamic range, elevated count-rate linearity, and strong detective quantum efficiency. For pump-probe schemes, they must also incorporate synchronized gating or photon-arrival-time discrimination. This timing capability becomes vital when an experiment needs to isolate a specific electron bunch embedded within a dense multibunch or hybrid filling pattern. Several current technologies showcase the strides being made to address these engineering demands. The GigaFRoST system, engineered at the Swiss Light Source, pioneered continuous image acquisition and data streaming at rates up to \qty{7.7}{\giga\bit\per\second} to a dedicated backend server, managing online preprocessing and data reduction on the fly\cite{bib-MOK2017}. While tailored for fast X-ray tomographic microscopy rather than direct time-resolved soft X-ray ptychography, its architecture serves as an instructive blueprint for the detector-to-computing pipelines required today. Meanwhile, Timepix4-based setups like TEMPUS deliver high-speed frame modes reaching $40000$ frames per second alongside event-driven detection capable of nanosecond-precision photon time-stamping\cite{bib-COR2024}. These tools are highly promising, though their deployment in soft X-ray ptychography depends heavily on sensor efficiency, noise performance, and maximum count rates in the lower energy regime. Encouragingly, detectors tuned specifically for soft X-rays are emerging. The EIGER-iLGAD detector, which merges a hybrid-pixel readout architecture with inverse low-gain avalanche diode sensors, has successfully demonstrated single-photon counting down to roughly \qty{550}{\electronvolt}, enabling ptychographic imaging at the Fe L$_3$ edge (\qty{707}{\electronvolt})\cite{bib-BAR2025}. This milestone is particularly relevant for magnetic imaging of 3d transition metals, whose primary $L_{\{2,3\}}$ absorption edges (Mn, Fe, Co, Ni) reside squarely within the soft X-ray regime. Nevertheless, bridging the gap between current speeds and the full temporal structure of DLSR sources will require further refinements in readout bandwidth, bunch-selective gating, and time-tagging accuracy. This soft-X-ray-optimized EIGER-iLGAD variant stands distinct from standard EIGER2 detectors, whose silicon sensors are optimized for much harder radiation.

The data rates spawned by these experiments will challenge the limits of acquisition electronics and computing infrastructure. A raw, uncompressed stream from a $2048\times2048$ pixel detector operating at \qty{1}{\kilo\hertz} with 16-bit depth generates about \qty{67}{\giga\bit\per\second} exclusive of metadata and network overheads. Data streams scaling to several tens of gigabits per second are therefore bound to become routine. Navigating this data deluge requires a tight integration of hardware and software: FPGA-driven preprocessing must handle online background subtraction, bad-pixel masking, and data reduction before passing the stream to GPU-accelerated clusters for rapid ptychographic reconstruction. Standard engines like the extended ptychographic iterative engine (ePIE), difference maps, and maximum-likelihood algorithms can be deployed within high-performance environments such as PyNX and PtychoShelves. Seamless detector-to-computing structures, echoing the GigaFRoST framework, offer a viable path forward\cite{bib-MOK2017,bib-BUT2025}.

\begin{acknowledgements}
We thank the Keysight company for providing us temporarily with the M8195A AWG and UXR0164B oscilloscope for the measurements, and Grégoire de Loubens for providing us with high-frequency signal generators for conducting preliminary measurements. We acknowledge support from the team of the Nanofab platform (Institut Néel) and of the Upstream Technological Platform PTA in Grenoble. We  thank Laura Alvaro-Gomez for preparing and providing some of the samples used in the experiments, and Nicolas Jaouen for useful discussions. A CC-BY public copyright license has been applied by the authors to the present document and will be applied to all subsequent versions up to the Author Accepted Manuscript arising from this submission, in accordance with the grant’s open access conditions \cite{bib-CC-BY}.
\end{acknowledgements}

\begin{funding}
This research received funding from France 2030 government investment plan managed by the French National Research Agency under grant reference SPINCHARAC ANR-22-EXSP-008, the French RENATECH network implemented at the Upstream Technological Platform in Grenoble PTA  (ANR-22-PEEL-0015), the French National Research Agency in the framework of the "Investissements d’avenir” program (ANR-15-IDEX-0002) and QuantAlps research federation FR2053 in Grenoble; Project PID2024-155385NB-C31 funded by MICIU/AEI/10.13039/501100011033 and by ERDF, EU and by Comunidad de Madrid via grant TEC-2024/TEC-380 (Mag4TIC-CM).
\end{funding}

\ConflictsOfInterest{There are no conflicts of interest.}

\DataAvailability{Data supporting the results can be made available by the authors upon reasonable request}

%\bibliographystyle{unsrt}
%\bibliographystyle{apsrev4-2}
%\bibliographystyle{iucr}
%\bibliography{Fruche8}

\end{document}